# Neutron relative effectiveness factors in Boron Neutron Capture Therapy: estimation of their values from the secondary charged particles and evaluation of weighted kerma factors for a standard tissue

M. Pedrosa-Rivera[1], J. Praena[1], C. Ruiz-Ruiz[2], M.J. Ruiz-Magaña[3] and I. Porras[1]

*[1]Departamento de Física Atómica, Molecular y Nuclear, Facultad de Ciencias, Universidad de Granada. 18071 Granada, Spain.*

*[2]Departamento de Bioquímica y Biología Molecular 3 e Inmunología, Facultad de Medicina, Universidad de Granada. 18016 Granada, Spain.*

*[3]Departamento de Biología Celular, Facultad de Ciencias, Universidad de Granada. 18071 Granada, Spain.*

## Abstract

The average Relative Biological effectiveness (RBE) factors for neutron irradiation in the context of a BNCT treatment are studied. This research considers the various interactions and secondary particles of each process and estimates the RBE based on the damage induced in tissues by all of these particles. A novel concept of estimating the biological dose by means of weighted kerma factors is introduced. These weighted kerma factors include the RBE of each energy deposition based on an RBE-LET relationship for secondary charged particles and can be directly incorporated in weighted dose calculations from Monte Carlo simulations. Furthermore, the dependence of the neutron weighting factor on neutron energy for standard soft tissue is discussed.

## 1. Introduction

The prediction of the biological damage produced by an external neutron irradiation is a complicated task. The main source of difficulty is that the ionization and DNA damage occurs indirectly, caused by secondary particles that are emitted from the neutron and photon interactions with the elements of tissue. In Boron Neutron Capture Therapy (BNCT), where beams of epithermal neutrons are used and they are thermalized in the tissues, the main interactions are: elastic scattering with hydrogen (producing protons of different energies up to the maximum neutron energy), nitrogen capture (producing 584 keV protons and 42 keV recoil carbon ions), hydrogen radiative capture (producing 2.224 MeV photons) and, where it is present, boron capture. This last capture produces, in 94% of the cases, a 1.47 MeV alpha particle and 0.84 MeV recoil $^7$Li, and in the other 6%, a 1.77 MeV alpha and 1.01 MeV $^7$Li. Other minor processes are the elastic scattering or reactions with other nuclei, like $^{18}$O, $^{12}$C or $^{14}$N. This causes the damage to be substantially dependent not only on the energy of the neutron, but also on the tissue components with which it interacts.

The ICRP Publication 60 [ICRP 2007], for the purpose of radiation protection, describes the neutron Relative Biological Effectiveness (RBE) as a constant value of 5 until $10^{-2}$ MeV, then it raised until a maximum value of 20 for 1 MeV neutrons and subsequently decreases back to the same constant value. These values were based on the limited available data and could be overestimated for radioprotection purposes, but when neutrons are used in radiotherapy, it is desirable to have more accuracy on the estimation of the biological damage for designing a treatment plan.

In BNCT the radiation dose delivered is described by means of four different components: *thermal dose*, $D_t$, from neutrons of less than 0.5 eV (where the main reaction is the capture by nitrogen); epithermal or *fast dose*, $D_f$, for neutrons with energies above 0.5 eV (where elastic scattering in hydrogen will be the main component); Both exclude , the *boron dose*, $D_B$ resulting from the primary reaction in BNCT, which is the main dose component at the tumor site; and finally, the *gamma dose*, $D_\gamma$, including prompt-gammas after captures and the gammas from the beam. These contributions, corresponding to different secondary particles which may differ in the linear energy transfer (LET), and hence, in their RBE, are weighted by constant factors (factors that do not include the dependence on dose, energy or tissue) providing the so-called *weighted or equivalent dose*, given by:

$$D_W = w_t D_t + w_f D_f + w_B D_B + w_\gamma D_\gamma. \qquad (1)$$

This weighted dose is used in treatment planning since many years ago, with the weighting factors reported, for example, by Coderre and Morris [Coderre and Morris 1999].. In particular, the same value is assumed for $w_t$ and $w_f$, what means in practice that the neutron effect is considered independent of the energy. The argument was that the proton from nitrogen capture (585 keV) and the mean energy of proton recoils of the fast neutrons from a reactor based source are similar. This may not be true for the new accelerator-based facilities and a dependence of $w_f$ on the spectrum is expected.

In addition to this, the use of $D_W$ as in Eq. (1) has been discussed and more realistic formalisms have appeared to describe the biological effect of BNCT by means of the *photon isoeffective dose* [Gonzalez and Santa Cruz 2012, Pedrosa-Rivera et al. 2020], which takes into account the linear-quadratic model (LQ) in the dose-effect relationships instead of a fixed RBE factor. However the requirement for a high number of experimental radiobiological data makes it difficult to impose them in the application on patients. Clearly there is a strong dependence on the tissue/cell line used as well as on the neutron beam used in the irradiation. Nevertheless, other factors such as the depth of the tumor, the boron compound used or the fractionation scheme have also a strong influence on the final RBE. Therefore, the need for a wide range of experimental data under different conditions is an accepted fact that is necessary for the improvement of BNCT treatment planning. However, a study of the general behavior of the biological effects can help where experimental data are lacking and give an approximate overview of the expected neutron damage during a BNCT treatment.

This paper attempts to shed some light on the neutron RBE overall dependency on energy by investigating theoretically the damage of each of the secondary particles emitted. Based on theoretical expressions of the neutron flux and the energy of the proton recoil, earlier work has explored the general behavior of the RBE depending on the neutron energy [Blue et al. 1993, Blue et al. 1995]. These works include the influence of protons from $^1$H(n,n')$^1$H in the first study, while the effect of $^{14}$N(n,p)$^{14}$C is simply a normalization, approach that is more precisely incorporated in the second work in 1995, which also uses a model by Fairchild and Bond [Fairchild and Bond 1985] for the RBE of protons. The approach presented in this paper has a similar objective, but applying more recent concepts of particle transport and dosimetry.

## 2. Method: Weighted-Kerma formalism for the biological dose

### 2.1 The kerma approximation for the absorbed dose

In reference BNCT Monte Carlo calculations [Goorley et al. 2002] the neutron absorbed dose rate $\dot{D}_n$ is calculated by integrating the product of neutron flux $\dot{\Phi}(E)$ (where $E$ denotes the neutron energy) and the neutron kerma factor $F_n(E)$, as:

$$\dot{D}_n = \int dE \, F_n(E) \, \dot{\Phi}(E) \,,$$

(2)

and the kerma factor is given by:

$$F_n(E) = \sum_{k,j} \sigma_{kj}(E) \, \frac{x_j}{A_j} N_A \, \epsilon_{jk}.$$

(3)

In this equation $\sigma_{kj}$ refers to the cross section of a process $k$ on the element $j$, $x_j$ and $A_j$ its fraction of mass and atomic mass, respectively, and $\epsilon_{jk}$ the energy delivered in the interaction ($N_A$ denotes the Avogadro's number).

Neutron kerma factors are described by Caswell [Caswell 1982] and a list of data for various tissues, based on cross section data from ENDF/B-VI [Young 1984], can be found in ICRU 63 [ICRU 2001].

This kerma approximation of the dose is valid when the mass element in which the dose is evaluated is in charged particle equilibrium i.e. the energy deposited outside the mass from interactions produced inside this element compensates for the energy deposited inside from interactions that occur outside it. Except at interfaces, this typically occurs within any homogeneous substance. Heavy charged particles are the dominant secondary particles produced by neutron interactions in BNCT (excluding photon production, which is treated independently). As a result, the prior assumption, that they deposit the energy locally, holds true for any point in the material except those located within a few microns of the interface with other media. This assumption was also used in the calculations conducted by Goorley [Goorley et al. 2002], which are widely regarded as an standard for BNCT absorbed dose estimations.

With this approach it is possible to separately estimate the thermal and fast neutron doses by limiting the energy ranges of the integrals as follow:

$$\dot{D}_t = \int_0^{0.5\,eV} dE\, F_n(E)\, \dot{\Phi}(E)\,, \tag{4}$$

and

$$\dot{D}_f = \int_{0.5\,eV}^{\infty} dE\, F_n(E)\, \dot{\Phi}(E)\,. \tag{5}$$

For the boron component, the boron kerma $F_n(E)$ is used. It can be evaluated with Eq.(3) but just using the interaction processes with $^{10}$B. This leads to:

$$\dot{D}_B = \int dE\, F_B(E)\, \dot{\Phi}(E)\,, \tag{6}$$

The gamma dose rate can be estimated also with photon kerma factors as in the work of Goorley et al.

The kerma factors can be calculated following previous calculations with the steps detailed in the work of Porras et al. [Porras et al. 2014], which includes expressions for calculating the contributions of the different secondary particles (for elastic collisions, for the emitted proton after capture and for the recoil after the capture). Results of this calculation, for a standard four components soft tissue (ICRU 33)[ICRU 46], but using real cross sections from ENDF database instead of the fitting functions approximations are shown in Figure 1.

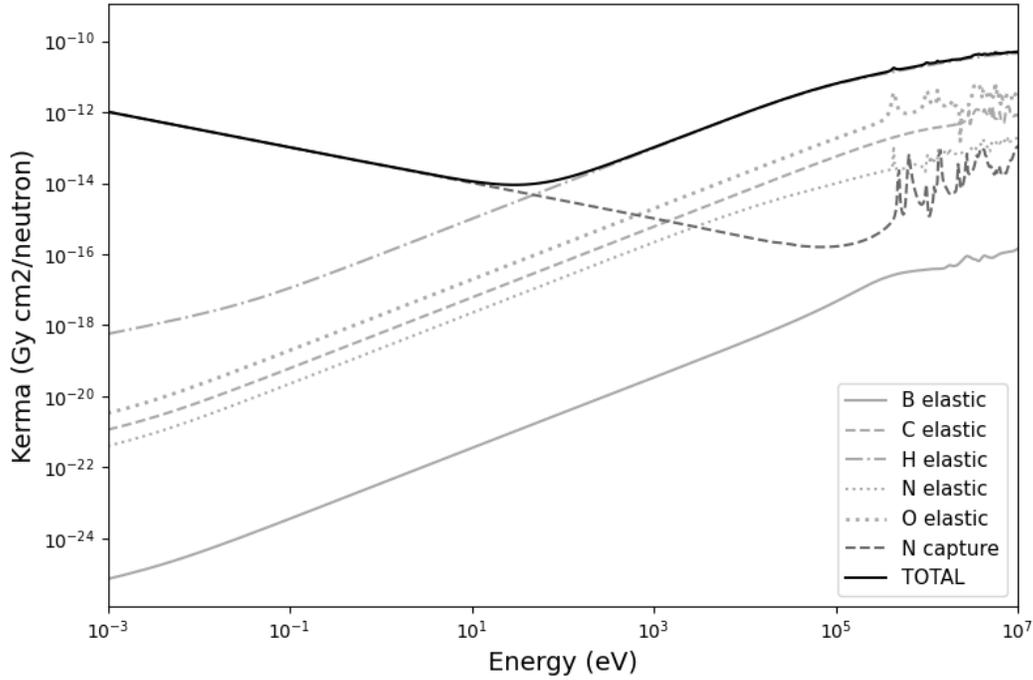

*Figure 1. Neutron kerma factors for ICRU 33 tissue and the individual element contributions.*

## 2.2. The weighted-kerma approximation for the biological dose

As the kerma factors are calculated by adding up the energy deposited by the different interactions, we propose using a weighting factor for each energy deposition before the summation, i.e.

$$F_n^W(E) = \sum_{j,k,q} \sigma_{kj}(E) \frac{x_j}{A_j} N_A \epsilon_{kj}^q \, RBE_q \,, \qquad (7)$$

where $q$ denotes the different secondary charged particles (can be more than one), produced in the interaction of type $k$ with the nucleus $j$. $\epsilon_{kj}^q$ is the energy of each particle $q$, and $RBE_q$ denotes its relative biological effectiveness.

Thus, if the biological impact of each secondary particle from each process is considered, the result should be more accurate than using a total weighting factor multiplied by each dose component. The integration of this weighted kerma factor with the neutron flux would directly produce an estimation of the weighted dose:

$$\dot{D}_n^W = \int dE \, F_n^W(E) \, \dot{\Phi}(E) \,, \tag{8}$$

Individual contributions of the different processes will also be analized in this work. For example, if we restrict the sum to a specific type of interaction (e.g. elastic collision). In Eq.(7) all processes are taken into account. For fast neutrons the dominating term accounts for the RBE of the recoil protons from the elastic scattering with hydrogen and for thermal ones that of the protons and carbon recoil from nitrogen capture, but we have considered all other possible minor processes of interaction, each with its individual RBE.

A boron weighting kerma factor $F_B^W(E_i)$ can be defined as in Eq. (7) but just taking into account the interactions with $^{10}$B:

$$F_B^W(E) = \sum_{k,q} \sigma_{k,10B}(E) \frac{x_{10B}}{10} N_A \epsilon_k^q \, RBE_q \,, \tag{9}$$

In this cases the processes $k$ are the two possible capture reactions (one leaving the lithium nuclei in its ground state (6%) and the other in the excited state). Particles $q$ in each case are the alpha particle and the lithium nucleus, with energies $\epsilon_k^q$ in each case. $RBE_q$ depends implicitly in $k$ as it corresponds to the RBE of the alpha particle and the lithium ion emitted in the neutron-boron reactions. The boron dose rate is just evaluated by:

$$\dot{D}_B^W = \int dE \, F_B^W(E) \, \dot{\Phi}(E) \,, \tag{10}$$

The values of the $RBE_q$ of each secondary particle, are the input data in the present formalism. They should be taken from experimental radiobiological data and may depend on the biological end-point chosen. However in section 3 we will use data from general LET-RBE relationships to predict the dependence of a total neutron RBE on the neutron energy.

## 2.3. Total neutron and boron RBE factors

After the effect of each secondary particle in each process is weighted, the total average RBE for neutrons as a function of the neutron energy $E$, can be calculated as follows:

$$\overline{RBE_n}(E) = \frac{F_n^W(E)}{F_n(E)}. \tag{11}$$

This provides an average of the RBE of the different secondary particles produced by a neutron of energy $E$. If $E < 0.5$ eV, this value can be called $\overline{RBE_t}(E)$ (thermal) and will be constant as the secondary particle as the kinetic energy of the neutron is negligible.

For higher energies, the value, which can be called $\overline{RBE_f}(E)$ (fast) will be energy-dependent and for an estimation of a global value for a particular facility it should be averaged with the neutron spectrum.

Similarly, the boron RBE can be estimated by:

$$\overline{RBE_B}(E) = \frac{F_B^W(E)}{F_B(E)}. \tag{12}$$

It must be remembered that photons have also a biological effect that depends mainly on the dose rate and photon energy. Since the objective of this work is the study of the effect of neutrons, the RBE value of the photons will be considered as the reference and therefore, for simplicity, will have the value of unity.

Given that they weight each dose/kerma component, these average RBE factors can be compared to the well-known weighting factors, $w_j$, if the end-point chosen is the same. Nonetheless, they have one major difference in that $\overline{RBE_f}(E)$ is dependent on the neutron energy, and for an estimation of a global value $w_f$ for a particular facility it should be averaged with the neutron spectrum.

The advantage of this formalism, is that once the weighted energy-dependent kerma factors are evaluated, it is very easy to include them in Monte Carlo simulations by introducing them as a table instead of the kerma factors in the transport calculation. They will include effectively the biological dose of the secondary particles in the calculation of the final weighted dose defined by Eqs. (8,10).

# 3. Results

## 3.1. Assumptions and approximations

In order to estimate the value of these average RBE for a specific tissue, the described process is applied in a 4-component tissue known as ICRU-33 tissue [ICRU 1992]. Before the application, there are a series of important factors that will influence the calculation. One is the RBE-LET data used and the other one is the LET calculation method.

The particle RBE-LET general behavior is known. Normally low LET corresponds to low RBE, with a value of 1, then there is an increase that reach maximum at the order of 100 keV/um, and then the RBE decreases with the increase of the LET. The specific values depend on the particle and the tissue and can be known by experimental data or modeling. For the example of application introduced in this work, the RBE-LET will be taken from the study of Barendsen [Barendsen 1994], that is a collection of experimental data corresponding to mammalian cells and different charged particles.

Specifically, the data used corresponds to single track lethal damage, expressed as STLD in the study [Barendsen 2001, Franken 2011], and shown in Figure 2. The reason for this decision is that these values correlate to the ratio of the alpha coefficients for charged particles and photons in the LQ model for this effect, i.e. the main effect for high-LET particles. With this end-point selected, then the $\overline{RBE_n}(E)$ and $\overline{RBE_B}(E)$ would be equal, respectively to the $W_j$-factors defined in [Pedrosa-Rivera 2020], which corresponds to the low-dose limit of the $w_j$-factors.

The values of Barensden of Figure 2 are fitted to a LET function, yielding the following expression:

$$RBE_q = \frac{1 + 0.005(2)\,LET + 0.00027(2)\,LET^2}{1 - 0.0146(4)LET + 0.000102\,(5)LET^2}. \tag{13}$$

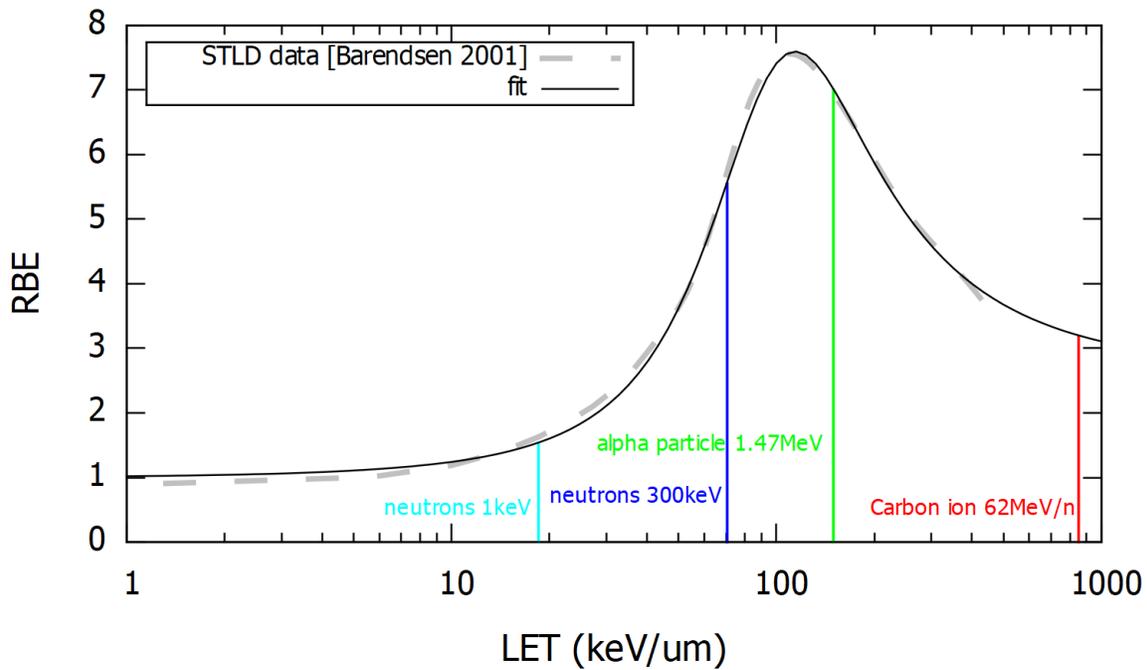

*Figure 2: Barendsen et al. (2001) data for mammalian cells and single track lethal damage and the fitting to the data of equation (13). It is marked by vertical lines some example values (see text in section 3.2): an epithermal neutron energy (1 keV), a fast one (300 keV), the main alpha particle emitted by the reaction between neutrons and $^{10}B$ (1.47 MeV)[ Jin 2022], and a Carbon ion beam of 62MeV/n at the center of the spread-out Bragg peak [Romano 2014]*

LET of the different secondary particles can be obtained from the known tables (NIST or SRIM), but as for low energetic particles a discrepancy is found, we have applied the continuous slowing down approximation (CSDA) to obtained the range of each particle, $R_{A,z}$ at each energy, $\varepsilon$, from the range of the protons, $R_{1,1}$ . Data from PSTAR at NIST [PSTAR web, ICRU 1993] data base was applied for energies above 1 keV while empirical data from Andersen and Ziegler [Andersen 1977] was used to fit low energies. Therefore, the range of each particle is calculated as:

$$R_{A,z}(\varepsilon) = \frac{A}{Z^2} R_{1,1}(\varepsilon/A), \qquad (14)$$

Then, the average LET of each particle is estimated with $\varepsilon$, the tranfered energy and $R_{A,z}$, the range in the media, as:

$$LET = \frac{\varepsilon}{R_{A,z}}. \qquad (15)$$

With these values, calculated easily for each secondary charged particle, their RBE can be estimated and introduced in Eqs. (7,9) for the evaluation of the weighted kerma factors $F_n^W(E)$ and $F_B^W(E)$. These values can be directly used as dose functions in Monte Carlo simulations. For this purpose we have calculated them for the standard ICRU 33 tissue. Values for a few energies are shown in Table 1 but a much larger table is given as an excel sheet available as supplementary information.

| Energy (eV) | $F_n^W(E)$ (Gy cm$^2$/neutron) | $F_B^W(E)$ (Gy cm$^2$/neutron) |
|---|---|---|
| 0.001 | 4,1797E-12 | 1,8876E-11 |
| 0.005 | 1,8705E-12 | 8,4472E-12 |
| 0.010 | 1,3227E-12 | 5,9733E-12 |
| 0.020 | 9,3533E-13 | 4,2237E-12 |
| 0.050 | 5,9135E-13 | 2,6702E-12 |
| 0.1 | 4,1832E-13 | 1,8885E-12 |
| 0.5 | 1,8723E-13 | 8,4491E-13 |
| 1.0 | 1,3249E-13 | 5,9733E-13 |
| 5.0 | 5,9947E-14 | 2,6694E-13 |
| 10 | 4,3378E-14 | 1,8867E-13 |
| 50 | 2,6378E-14 | 8,4186E-14 |
| 100 | 2,8571E-14 | 5,9448E-14 |
| 500 | 8,2473E-14 | 2,6441E-14 |
| 1000 | 1,5687E-13 | 1,8630E-14 |
| 5000 | 8,0546E-13 | 8,2401E-15 |
| 10000 | 1,7369E-12 | 5,8071E-15 |
| 50000 | 1,1564E-11 | 2,6880E-15 |
| 100000 | 2,5820E-11 | 2,0438E-15 |
| 500000 | 8,7727E-11 | 1,0666E-15 |
| 1000000 | 1,0467E-10 | 4,3728E-16 |
| 2000000 | 8,1194E-11 | 8,5118E-16 |
| 10000000 | 7,6501E-11 | 7,5465E-16 |

*Table 1: Values of the Weighted Kerma Factors $F_n^W(E)$ for a 4-component soft tissue (ICRU 33, 76,2 % oxygen, 11,1 % carbon, 10,1 % hydrogen and 2,6 % nitrogen) and of the boron Weighted Kerma $F_B^W(E)$ for a 10 ppm of $^{10}B$ concentration.*

In figure 3, a plot of these $F_n^W(E)$ values and their contributions from capture and scattering events is displayed. Please note that the values expressed in Gy in this table represent weighted values, which are sometimes denoted as Gy-eq in the BNCT literature. They are also compared to the (non-weighted) kerma factors. It can be noticed the minimum in the epithermal region (the most adequate for BNCT of deep seated tumors), where the roles of both types of process are switched.

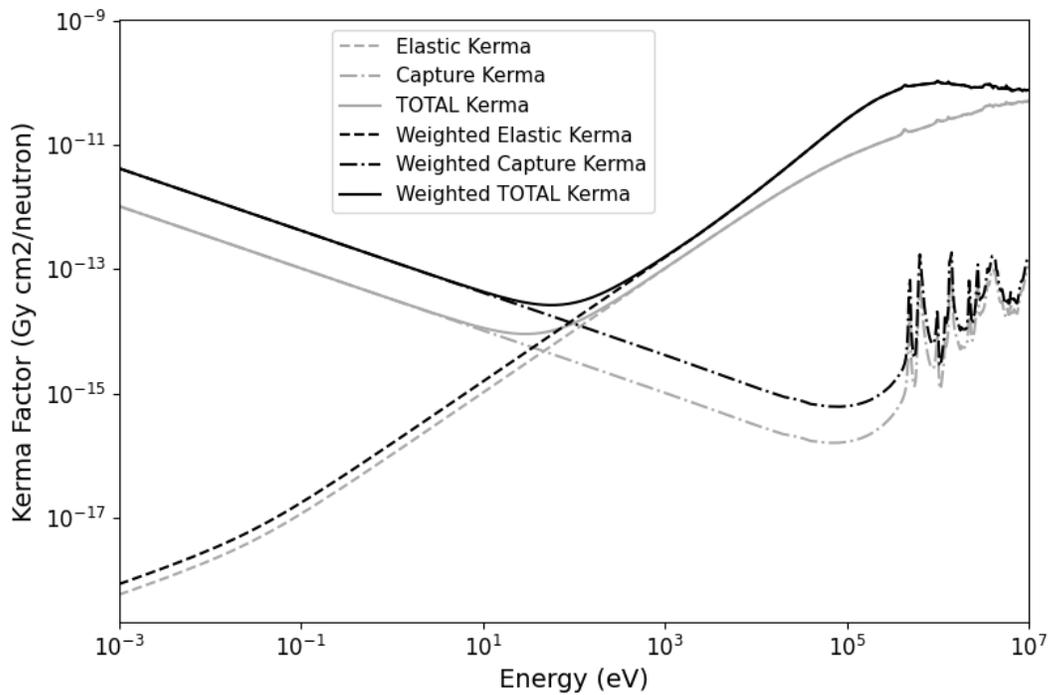

*Figure 3: Elastic, capture and total kerma factors (grey) and their corresponding weighted kerma factors (black) for ICRU-33 4-components tissue.*

In figure 4, a plot of the $F_B^W(E)$ are displayed and compared to the boron kerma factor $F_B(E)$. Both are parallel except at high neutron energies, as the kinetic energy of the secondary particles are dominated by the $Q$ value of the reactions.

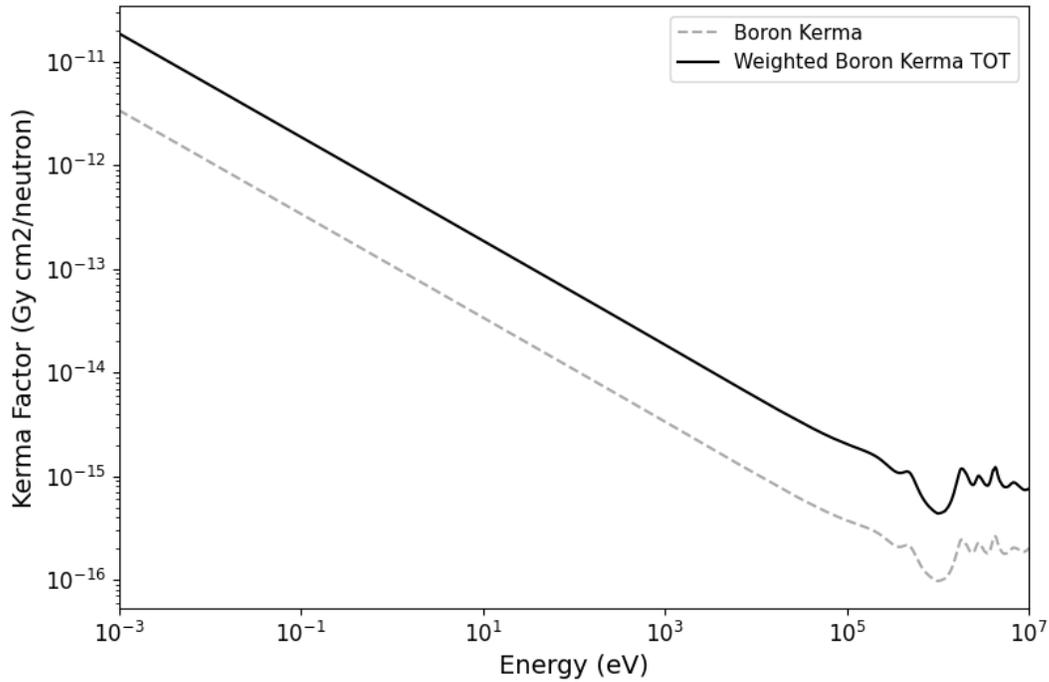

*Figure 4: Boron kerma factor (grey) and weighted boron kerma factor (black) for a tissue with a mass fraction of boron of $10^{-5}$.*

### 3.2. Neutron RBE for ICRU-33 tissue

With the values of the neutron weighted kerma factors and kerma factors calculated, their ratio represent the global RBE $\overline{(RBE_n}(E))$ of the neutrons of each energy, as given by Eq. (11). They are displayed in Figure 5. It is worth to note how there is a clear minimum at epithermal energies, where accelerator-based neutron sources have a peak in the spectrum.

Figure 5 also presents previously reported data on RBE estimations from Blue et al., which exhibit a similar trend, though differing in absolute values due to variations in normalization methods. One dataset has been normalized using the current RBE factor of 3.2 applied in BNCT, while the other is normalized based on Fairchild's data [Fairchild, 1985]. All referenced datasets utilize the same Barendsen et al. [Barendsen 2001] data corresponding to 10% cell survival.

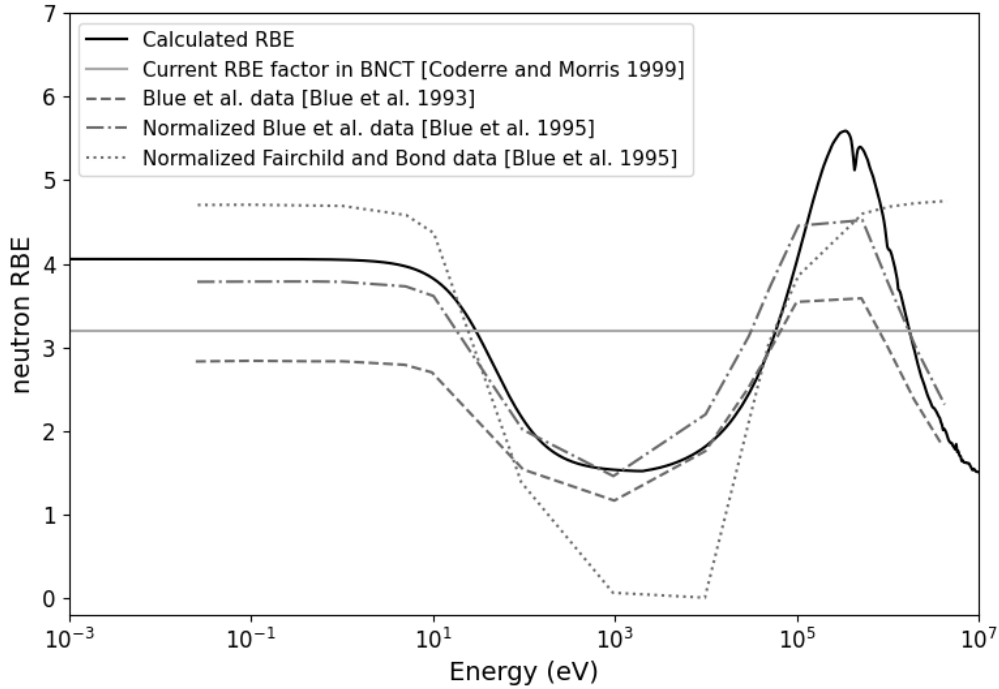

*Figure 5: $\overline{RBE_n}(E)$ for ICRU-33 tissue (black line) as a function of neutron energy. The dotted lines correspond to the data shown in [Blue et al. 1993, Blue et al. 1995]. The horizontal grey line indicates the current 3.2 weighting factor in BNCT, for all neutron energies.*

For this biological end-point, and as mentioned before, these $\overline{RBE_n}(E)$ are equal to an effective alpha-coefficient ratio between the neutron and photons and can provide an estimation of the $W_n$ factors defined in [Pedrosa-Rivera et al. 2020]. These factors represent the ratio of the alpha coefficients for neutron and photon doses, which corresponds to the maximum value of the standard weighting factor at the low-dose limit. The $W_t$ factor coincides to the constant value of $\overline{RBE_n}(E)$ for energies below 0.5 eV, which for this standard tissue, is equal to 4.06. For neutrons above 0.5 eV, as the $\overline{RBE_n}(E)$ is energy-dependent, the $W_f$ factor should be obtained by the average of $\overline{RBE_n}(E)$ over the non-thermal spectrum:

$$W_f = \frac{\int_{0.5\,eV}^{\infty} dE\ \overline{RBE_n}(E)\ \Phi(E)}{\int_{0.5\,eV}^{\infty} dE\ \Phi(E)} \qquad (16)$$

Clearly, for a pure epithermal spectrum for which most neutrons are below 10 keV, this value should be smaller than the 3.2 value used as the weighting factor of the fast dose. In table 2 we provide values of $W_f$ for some specific energies (assuming different monoenergetic neutrons). Moreover, the biological effect of the epithermal neutrons will vary significantly depending on the beam spectrum, with particular importance placed on whether the beam includes neutrons around 300 keV, where the maximum RBE is observed.

| **Energy:** | Less than 1 eV | | 10 eV | 100 eV | 1 keV | 10 keV | 100 keV | 500 keV | 1 MeV |
|---|---|---|---|---|---|---|---|---|---|
| $W_n$: | $W_t$: 4.06 | $W_f$: | 3.83 | 2.12 | 1.54 | 1.82 | 4.02 | 5.40 | 4.19 |
| $W_B$: | 5.52 | | 5.52 | 5.52 | 5.52 | 5.52 | 5.46 | 5.08 | 4.49 |

*Table 2: Estimated values of the weighting factors, $W_i$ [Pedrosa-Rivera et al. 2020], for ICRU-33 tissue for average mammalian cells from the RBE of the secondary charged particles based on RBE-LET STDL Barendsen et al. data [Barendsen2001] and with a mass fraction of boron of $10^{-5}$.*

In figure 2 the values of the RBE obtained for two representative energies (1 keV and 300 keV) are illustrated by means of vertical lines. Also, for comparison, the corresponding RBE for the same biological endpoint (single track lethal damage) for the main alpha particle produced by the boron neutron reaction [Jin 2022] and for a carbon ion from the literature [Romano 2014] are also displayed. From this comparison it can be highlighted the importance of avoiding neutrons for BNCT above 100 keV for sparing normal tissues, as their RBE in that energy range can be superior to those of C ions.

### 3.3. Boron RBE for ICRU-33 tissue

$\overline{RBE_B}(E)$ is approximately constant, and it then can be assumed to be equal to $W_B$, which is found to be 5.52, as displayed in table 2.

However, we should take into account that this is a general estimation from mammalian cells ratio of the alpha coefficients between boron dose and photon dose, so it corresponds to the maximum value of the standard weighting factor $w_B$ in the low dose-limit.

Moreover, this value is found from data from uniform irradiations with alpha particles. It is well know that $w_B$, also called CBE (Compound Biological Effectiveness), is specific on the compound as the microdistribution inside the cell may affect strongly the biological effect. This should be measured for each compound by means of radiobiology experiments, and the value reported here is just a general value that could be significant if the distribution of the boron is uniform inside the cell. In addition to this, it represents an average of measurements in different cell lines and it should be measured at the most representative cell lines for a given treatment.

### 3.4. Use of the weighted kerma factors for the evaluation of the photon isoeffective dose in Monte Carlo simulations

Monte Carlo simulation codes are able to evaluate, for a given region of tissue, the neutron fluence per unit energy and initial neutron $\widehat{\Phi}(E)$, from which the weighted dose can be estimated multiplying by the initial neutron beam intensity $I(t)$ and the weighted kerma factors, and integrating at all energies and irradiation time:

$$D_n^W = \int dt\, I(t) \int dE\, F_n^W(E)\, \widehat{\Phi}(E)\,, \tag{17}$$

and for the boron weighted dose:

$$D_B^W = \int dt\, I(t) \int dE\, F_B^W(E)\, \widehat{\Phi}(E)\,. \tag{18}$$

As the dose is already weighted at all energies, there is no need to separate the thermal and fast components in Eq. (17).

Finally, we can estimate the photon isoeffective dose by using the approximation of [Pedrosa-Rivera et al. 2020 a] to the formalism defined by [Gonzalez and Santa Cruz 2022]:

$$D_n^W + D_B^W + D_\gamma + \frac{D_\gamma^{\,2}}{\alpha_\gamma/\beta_\gamma} = D_{isoE} + \frac{D_{isoE}^{\,2}}{\alpha_\gamma/\beta_\gamma}\,, \tag{19}$$

and then $D_{isoE}$ would be given by

$$D_{isoE} = \frac{\alpha_\gamma/\beta_\gamma}{2}\left[-1 + \sqrt{1 + 4\,\frac{D_n^W + D_B^W + D_\gamma + \dfrac{D_\gamma^{\,2}}{\alpha_\gamma/\beta_\gamma}}{\alpha_\gamma/\beta_\gamma}}\right] \tag{20}$$

The values of $\alpha_\gamma/\beta_\gamma$ corresponds to the well-known radiobiological factor $\alpha/\beta$ used in photon radiotherapy for estimating the biological effective dose (BED) in conventional radiotherapy. Typically a value of 10 is assumed for early-responding tissues and tumors, and 3 for late-responding tissues (normal tissue) [Fowler 2024].

## Discussion and conclusions

The weighted kerma factors evaluated here are a similar concept as the standard kerma factors but including the information of the biological effect in general mammalian cells. They can be used for a gross estimation of the photon isoeffective dose, but particularities of the tissues, if known, should be considered. For example, if the effect of charged particles in a particular tissue is measured as a function of their LET, it can be introduced easily as the input $RBE_q$ data in this method.

The results shown for ICRU-33 tissue refer to an approximate and standard tissue calculation, so they reflect the generic behaviour of neutron damage with respect to energy. This general behaviour allows certain conclusions to be drawn and suggests future work to be carried out. However, it should be noted that the specific absolute values of $\overline{RBE_n}(E)$ as well as the weighted Kerma factors will be affected according to the tissue and endpoint experimental data used. However, the steps described in this work can be applied in the same way, but based on other experimental data.

The model is fed by experimental data, allowing its refinement and increased specificity for particular tissue types as more data become available. In the presented case, the results are derived from experimental data provided by Barendsen, pertaining to generic mammalian tissue. However, the model could be adapted to specific tissues if sufficient data exist for varying neutron energies or LET values. The endpoint considered here is the single-track lethal damage, quantified by the ratio of the alpha coefficients in the linear-quadratic model. Nonetheless, the model could be extended to incorporate other endpoints if relevant experimental data were available. For instance, data on potential lethal damage (PLD), also derive from Barendsen's work [Franken 2011], could be integrated in the same way. To expand the dataset, it would be advantageous to conduct *in-vitro* and *in-vivo* irradiation experiments with different particle beams, as in [Pedrosa-Rivera et al., 2020b; Mason et al., 2011, Krishnan 1977, Masunaga 2014, Kusumoto 2019], or to utilize existing databases like that of [Friedrich et al., 2021]. The model is also constrained by the limited availability of experimental data in areas beyond tissue types and endpoints. For instance, potential synergies between different contributions are not accounted for, nor is the impact of dose rate on photon interactions. Incorporating these effects would require extensive experimental data encompassing the entire energy spectrum and varying dose rates.

One of the most attractive potential applications of these newly introduced factors is that they can be used directly in the Monte Carlo simulations for dosimetry. With this, the biological dose can be estimated directly from the simulations, and it will include the effect at all energies and interactions realistically.

With the weighted kerma factors it is also possible to obtain the average neutron RBE for each neutron energy. Blue et al. [Blue et al. 1993, Blue et al. 1995] took a similar approach, calculating the energy dependence of the RBE of neutrons and then normalizing it to the neutron beam of the Brookhaven Medical Research Reactor. Blue et al. obtained an RBE-energy dependence that follows the same form as the result of this work at 10% survival, with a significant minimum at 0.1 keV. Both the results of Blue et al. and presented ones reveal the same conclusion: it is not accurate to assume a constant value for the neutron biological effect, because the RBE-energy relationship is not constant for all energies. As a result, using the same RBE value for thermal neutrons and epithermal neutrons is not realistic, as it is done in BNCT currently. Even if the $\overline{RBE_n}(E_i)$ in this work were computed with some approximations, it is a superior technique to assess the theoretical dependence of the biological effect with energy since it incorporates information about the individual effect of the primary secondary particles in a BNCT irradiation.

Connections with previous formalisms can be made. Since they correspond to the effect of the linear parameter in the LQ model, $\overline{RBE_n}(E_i)$ produced from the STLD data can directly linked to the new weighting factors, $W_i$, introduced in the evaluation of the photon iso-effective dose [Pedrosa-Rivera et al. 2020 a].

The weighted kerma factors as well as the presented approach to obtain the neutron RBEs can help when using the neutron field and when choosing the conditions during a BNCT irradiation. The study also encourages to more irradiation experiments necessary for clarifying the RBE-LET dependence for different particles, the neutron RBE at low neutron energy for different tissues and the epithermal neutron RBE of each facility.

## Acknowledgements

This work was partially supported by Spanish Instituto de Salud Carlos III (DTS22/00147), the Fundación Científica de la Asociación Española Contra el Cáncer (INNOV223579PORR), La Caixa Foundation (CC21-10047) and the Spanish Ministerio de Ciencia e Innovación (PID2020-117969RBI00). We also acknowledge financial support from the donors of the University Chair Neutrons for Medicine: Fundación ACS, Asociación Capitán Antonio, La Kuadrilla de Iznalloz and Costaleros Contra el Cáncer.

## References

Andersen H. H. and Ziegler J. F. (1977). Hydrogen: Stopping Powers and Ranges in All Elements. Vol. 3 of The Stopping and Ranges of Ions in Matter (Pergamon Press, Elmsford, New York). ISBN 0-08-021605-6

Barendsen G. W. (1994). The relationships between RBE and LET for different types of lethal damage in mammalian cells: biophysical and molecular mechanisms. Radiation Research 139(3), 257-270. https://doi.org/10.2307/3578823

Barendsen G. W., Van Bree C., Franken, N. A. (2001). Importance of cell proliferative state and potentially lethal damage repair on radiation effectiveness: implications for combined tumor treatments. International Journal of Oncology 19(2), 247-256. https://doi.org/10.3892/ijo.19.2.247

Blue T. E., Gupta N., Woollard, J. E. (1993). A calculation of the energy dependence of the RBE of neutrons. Physics in Medicine & Biology 38(12), 1693. DOI 10.1088/0031-9155/38/12/002

Blue T. E., Woollard J. E., Gupta N., Greskovich Jr J. F. (1995). An expression for the RBE of neutrons as a function of neutron energy. Physics in Medicine & Biology 40(5), 757. DOI 10.1088/0031-9155/40/5/004

Caswell R. S., Coyne J. J., Randolph M. L. (1982). Kerma factors of elements and compounds for neutron energies below 30 MeV. The International Journal of Applied Radiation and Isotopes 33(11), 1227-1262. https://doi.org/10.1016/0020-708X(82)90246-0

Coderre J. A., et al. (1993). Derivations of relative biological effectiveness for the high-LET radiations produced during boron neutron capture irradiations of the 9L rat gliosarcoma in vitro and in vivo. International Journal of Radiation Oncology, Biology, Physics 27(5), 1121-1129. https://doi.org/10.1016/0360-3016(93)90533-2

Coderre J. A., & Morris G. M. (1999). The radiation biology of boron neutron capture therapy. Radiation research, 151(1), 1-18. https://doi.org/10.2307/3579742

Fairchild R. G., Bond V. P. (1985). Current status of 10B-neutron capture therapy: enhancement of tumor dose via beam filtration and dose rate, and the effects of these parameters on minimum boron content: a theoretical evaluation. International Journal of Radiation Oncology* Biology* Physics 11(4), 831-840. https://doi.org/10.1016/0360-3016(85)90318-9

Fowler J.F. (2024). Radiation Biologically Effective Dose (BED) Calculator. https://www.mdcalc.com/calc/10111/radiation-biologically-effective-dose-bed-calculator. Checked on 29/9/2024.

Franken N. A., et al. (2011). Comparison of RBE values of high-LET α-particles for the induction of DNA-DSBs, chromosome aberrations and cell reproductive death. Radiation Oncology 6(1), 64. https://doi.org/10.1186/1748-717X-6-64

Friedrich T., Pfuhl T., Scholz M. (2021). Update of the particle irradiation data ensemble (PIDE) for cell survival. J Radiat Res, 62 (4), pp. 645-655 (2021). https://doi.org/10.1093/jrr/rrab034

González S. J., Santa-Cruz G. A. (2012). The photon-isoeffective dose in boron neutron capture therapy. Radiation Research 178, 609-621. https://doi.org/10.1667/RR2944.1

Goorley J. T., Kiger W. S., Zamenhof R. G. (2002). Reference dosimetry calculations for neutron capture therapy with comparison of analytical and voxel models. Medical Physics 29(2), 145-156. https://doi.org/10.1118/1.1428758

ICRP publication 60 (2007). Ann ICRP, 37(2.4), 2.

International Commission on Radiation Units and Measurements report 46, Bethesda M.D. (1992). Photon, electron, proton, and neutron interaction data for body tissues. International Commission on Radiation Units and Measurements,

International Commission on Radiation Units and Measurements report 49 (1993). Stopping Powers and Ranges for Protons and Alpha Particles.

International Commission on Radiation Units and Measurements report 63 (2001). Nuclear Data for Neutron and Proton Radiotherapy and for Radiation Protection. https://doi.org/10.1118/1.1369116

Jin W. H., Seldon C., Butkus M., Sauerwein W. Giap H. B. (2022). A review of boron neutron capture therapy: Its history and current challenges. *International journal of particle therapy*, *9*(1), 71-82. https://doi.org/10.14338/IJPT-22-00002.1

Krishnan D., Gopakumar K., Bhandari N. S., Madhvanath U. (1977). Response of aqueous coumarin solution to high LET radiation from 10B (n, α) 7Li reaction. *Radiation Effects*, *34*(4), 203-207. https://doi.org/10.1080/00337577708233148

Kusumoto T., Ogawara R. (2019). Radiation chemical yield of hydroxyl radicals for accelerator-based boron neutron capture therapy: Dose assessment of 10B (n, α) 7Li reaction using Coumarin-3-Carboxilic solution. *Radiation research*, *191*(5), 460-465. https://doi.org/10.1667/RR15318.1

Mason A. J., Giusti V., Green S., af Rosenschöld P. M., Beynon T. D., Hopewell J. W.(2011). Interaction between the biological effects of high-and low-LET radiation dose components in a mixed field exposure. International Journal of Radiation Biology 87, 1162-1172. https://doi.org/10.3109/09553002.2011.624154

Masunaga S. I. et al. (2014). The dependency of compound biological effectiveness factors on the type and the concentration of administered neutron capture agents in boron neutron capture therapy. *Springerplus*, *3*, 1-11. http://www.springerplus.com/content/3/1/128

NIST Standard Reference Database 124, https://physics.nist.gov/PhysRefData/Star/Text/PSTAR-t.html.

Pedrosa-Rivera M., Praena J., Porras I., Ruiz-Magaña M. J., Ruiz-Ruiz C. (2020 a). A simple approximation for the evaluation of the photon iso-effective dose in Boron Neutron Capture Therapy based on dose-independent weighting factors. Applied Radiation and Isotopes, 157, 109018. https://doi.org/10.1016/j.apradiso.2019.109018

Pedrosa-Rivera, M. et al (2020 b). Thermal Neutron Relative Biological Effectiveness Factors for Boron Neutron Capture Therapy from In Vitro Irradiations. Cells, 9(10), 2144. https://doi.org/10.3390/cells9102144

Porras I., Sabaté-Gilarte M., Praena J., Quesada J. M., Esquinas P. L. (2014). 33S for neutron capture therapy: nuclear data for Monte Carlo calculations. Nuclear Data Sheets 120, 246-249. https://doi.org/10.1016/j.nds.2014.07.058

Romano F., et al. (2014). A Monte Carlo study for the calculation of the average linear energy transfer (LET) distributions for a clinical proton beam line and a radiobiological carbon ion beam line. *Physics in Medicine & Biology*, *59*(12), 2863. DOI 10.1088/0031-9155/59/12/2863

Young A.G., (1984)ENDF/B-VI MAT 7925,